\documentstyle[epsfig]{mn}
\def\msun{\mbox{M$_{\odot}$}}
\def\mvir{\mbox{M$_{\rm v}$}}
\def\ms{\mbox{M$_{\rm s}$}}
\def\md{\mbox{M$_{\rm d}$}}
\def\vm{\mbox{V$_{\rm max}$}}
\def\fd{\mbox{f$_{\rm d}$}}
\def\fb{\mbox{f$_{\rm b}$}}
\def\fg{\mbox{f$_{\rm g}$}}
\def\rs{\mbox{r$_{\rm s}$}}
\def\infrate{\mbox{$\dot{\Sigma}_{\rm g}$}}
\def\Sg{\mbox{$\Sigma_{\rm g}$}}
\def\td{\mbox{t$_{\rm d}$}}
\def\sfr{\mbox{$\dot{\Sigma}_{\rm s}$}}
\def\ssd{\mbox{$\Sigma_{\rm s}$}}
\def\ssg{\mbox{$\Sigma_{\rm g}$}}
\def\kms{\mbox{km s$^{-1}$}}
\def\grtsim{{_ >\atop{^\sim}}}

\title[Cosmological solar neighbourhood star formation history]
{A cosmological study of the
star formation history in the solar neighbourhood}
\author[X. Hern\'{a}ndez, V. Avila-Reese and C. Firmani]
{X. Hern\'{a}ndez $^{1}$, Vladimir Avila-Reese$^{1}$ and Claudio Firmani$^{1,2}$\\
$^1$ Instituto de Astronom\'\i a, Universidad Nacional Aut\'onoma de
M\'exico, A.P. 70-264, 04510 M\'exico, D.F. \\
$^2$ Osservatorio Astronomico di Brera, via E. Bianchi 46, 23807 Merate (LC), Italy\\
}

\date{\today}

\begin{document}
\maketitle



\begin{abstract}

We use a cosmological galactic evolutionary approach to model
the Milky Way. A detailed treatment of the mass aggregation and dynamical history
of the growing dark halo is included, together with a self consistent physical
treatment for the star formation processes within the growing galactic disc.
This allows us to calculate the temporal evolution of star and gas surface densities
at all galactic radii, in particular, the star formation history (SFH) at the solar
radius. A large range of cosmological mass aggregation histories (MAHs) is 
capable of producing a galaxy with the present day properties of the Milky Way. 
The resulting SFHs for the solar neighbourhood bracket the available observational
data for this feature, the most probable MAH yielding the optimal comparison 
with these observations. We also find that the 
rotation curve for our Galaxy implies the presence of a constant density core in 
its dark matter halo.

\end{abstract}


\section{Introduction}

The star formation history (SFH) of galaxies is a key ingredient
in describing their evolution. From an observational point of view, the 
SFH can be studied by measuring the integrated light of the galaxy 
population as a whole or of a homogeneous sample of galaxies at different 
redshifts, at wavelengths associated with massive star formation (e.g., 
Madau, Pozzetti \& Dickinson 1998; Lilly et al. 1998).
An alternative method, which avoids the difficulties in selecting a homogeneous
sample of galaxies, determines the SFH of a specific galaxy using multi-band
photometry and chemical abundances combined with population synthesis models. 
In the case of the Milky Way (MW), the large amount of available data 
allows to qualitatively constrict the input SFHs of these methods
with some accuracy (e.g., Prantzos \& Aubert 1995; Chiappini, Matteucci, 
\& Gratton 1997; Carigi 1997; Boissier \& Prantzos 1999).   

On the same line of multi-band photometry, but with higher precision,
the SFH of a system can be obtained through the study of its resolved 
stellar population. This allows to construct colour-magnitude diagrams, 
which contain detailed information on the SFH behind
such diagrams. Inverting the problem however, is not trivial, with
the  answer depending sensitively on the assumed metallicity of the
object being studied at every time, i.e., the enrichment history.
The accuracy of these inferences being also crucially determined by the 
depth and error level of the observations, these requirements have 
so far severely constrained the applicability of such direct inferences 
to only a handful of astrophysical systems.
Still, such methods have recently been applied to systems where the
metallicity has been independently determined, and where high quality data exist.
Examples being the works of  Chiosi et al. (1989), 
Aparicio et al. (1990) and Mould et al. (1997) using Magellanic
and local star clusters, and Mighell \& Butcher (1992), Smecker-Hane et al. (1994), 
Tolstoy \& Saha (1996), Aparicio \& Gallart (1995), Mighell (1997) 
and Hernandez et al. (2000a) using local dSph's.

With the coming of high quality photometric data from the 
Hipparcos satellite, colour magnitude diagram inversion methods using
rigorous statistical analysis (Tolstoy \& Saha 1996; Hern\'andez et al.
1999 and more references therein) can be applied to the solar neighbourhood SFH. 
For example, the advanced Bayesian
analysis introduced by Hern\'andez et al. (2000b), allows the  
recovery of the underlying SFH without the need of assuming
any {\it a priori} structure or condition on the SFH. 
With the Hipparcos catalogue the time 
resolution of this technique is $\sim 0.05$ Gyr, however, the age 
range which it allows to explore is small (the last 3 Gyr); this limitation
is related to the completeness of the Hipparcos solar neighbourhood
sample. With a standard parametric maximization technique, Bertelli \& Nasi (2001)
were able to recover the SFH along the whole life of the 
solar neighbourhood, at the expense of a low time resolution. 

The high quality SFH which can be inferred locally
allows the unique opportunity of studying directly the evolution of an
individual cosmological object (the Galaxy). This avoids the problems inherent 
to high redshift studies, such as the uncertainties in relating a population 
of poorly understood and barely resolved objects to a particular class of 
present day systems.

Once the SFH of an observed system is known, the goal is to understand 
its physical evolution. From a theoretical point of view, the challenge 
is to link the SFH to the structure, dynamics and hydrodynamics of modeled 
galaxies. Examples of galaxy evolutionary studies in a cosmological 
context are: White \& Frenk (1991), Lacey \& Silk (1991), Kauffmann, 
White \& Guiderdoni 1993, Cole et al. (1994), Baugh, Cole \& Frenk (1997), 
van den Bosch (1998), Somerville \& Primack (1999), Buchalter, Jim\'{e}nez 
\& Kamionkowski (2001) and Kauffmann, Charlot \& Balogh (2001). A potential 
shortcoming of many of the above approaches remains the use of somewhat 
{\it ad hoc} or empirical schemes for calculating the star formation (SF) 
in the forming discs.

We improve on this last point by including a physically self consistent 
scheme of calculating the SF process, where the SF in the disc is induced by 
gravitational instabilities and is self-regulated by an energy balance 
in the ISM (Firmani \& Tutukov 1992, 1994; Firmani, Hern\'andez, \& 
Gallagher 1996). Combining the above with a cosmological evolutionary approach 
(Avila-Reese, Firmani \& Hernandez 1998 and Firmani \& Avila-Reese 2000) yields
a powerful tool for exploring the consequences of any particular cosmology
in terms of a detailed SFH (Avila-Reese \& Firmani 2001). 

In this work we make use of this evolutionary approach to infer the
SFH of the MW. The cosmological scenario used allows to constrain the SF history of the 
solar neighbourhood to a narrow range of possibilities, and to identify the most likely
of them. Comparison with the available direct observational inferences shows that the
local SFH has in fact been very close to the maximum likelihood solution we find.
Along similar lines, 
Hern\'andez \& Ferrara (2001) have studied the link between the cosmological 
build up of the Milky Way and the IMF of population III stars, by analyzing the
metallicity distribution of extremely metal poor Galactic halo stars.

In \S 2 a brief description of the method is presented,  
emphasizing on the physics of the SF. In \S 3 we apply this method to
model a MW galaxy. The SFH at the solar radius is 
presented and compared with observations in \S 4. Section 4.1  
is devoted to exploring the sensitivity of the predicted SFH to the 
input parameters which define our MW model; these parameters were settled 
from observational constraints which are subject to uncertainties. In the 
discussion we compare our results with those of chemo-spectrophotometric
models (\S 5.1), and we discuss the possibility of an intermittent
SF, as well as other physical mechanisms capable of triggering SF (\S 5.2). 
Our conclusions are presented in \S 6. Throughout this work we assume 
the currently favored cosmology, a flat universe with non-vanishing
vacuum energy ($\Omega_{m}=0.3, \Omega_{\Lambda}=0.7,
H_{0}=65 kms^{-1} Mpc ^{-1}, \sigma_{8}=0.9$).

\section{The Model}

The overall method of disc galaxy evolution used here was presented in  
Firmani \& Avila-Reese (2000, henceforth FA00) and 
Avila-Reese \& Firmani (2000, henceforth AF00) (see also 
Firmani, Avila-Reese \& Hern\'andez 1997). Galactic discs form inside-out 
within a growing CDM halo, as the baryonic content
of the growing halo is incorporated into the disc. The evolution of 
the halo is set by its mass aggregation history (MAH), 
determined by an extended Press-Schechter formalism.
For a given present-day total
virial mass \mvir, a statistical set of MAHs is calculated from 
the initial Gaussian density fluctuation field once the 
cosmology and power spectrum are fixed (Lacey \& Cole 1993; Avila-Reese et al. 1998). 
The behaviour of the growth of the halo in mass as a function of redshift can be seen 
in Fig. 1b, which shows a sample of five realizations of possible MAHs.
Fig. 1a gives the average over a statistical sample of $2\times 10^4$ Monte Carlo
realizations for a $2.8\times 10^{12}$ \msun\ halo. Although the ``average MAH'' 
clearly smoothes over the discontinuous behaviour of any particular MAH, it 
provides the simplified version for the most probable mass aggregation process of 
a large sample of galaxies. Further, the discontinuous behaviour seen in 
the individual MAHs of Fig. 1b, as reflected in gas accretion, will be 
naturally smoothed over by the gas virialization and cooling processes in the 
growing galactic halo.

The density profile of the virialized
part of the growing halo is calculated with a generalization of the
secondary infall model, based on spherical symmetry and adiabatic 
invariance, but allowing for non-radial orbits with fixed pericentre to apocentre
distance ratios $\epsilon$ (Avila-Reese et al. 
1998). The halo density profile we obtain depends on the MAH and $\epsilon$,
with the average MAH and $\epsilon \approx 0.2$ giving density profiles 
very close to the Navarro, Frenk \& White (1997, NFW) profile 
(Avila-Reese et al. 1998, 1999). 

A fraction \fd\
of the mass of each collapsing spherical shell is transferred 
in a virialization time into a central disc gas layer.
Assuming angular momentum conservation, a given gas element of 
the shell falls into the equatorial plane at the position where 
it reaches centrifugal equilibrium. Each shell is assumed to be initially in
solid body rotation, with a specific angular momentum given by:
\begin{equation}  
j_{sh}(t)=\lambda \frac{{G\mvir(t)^{5/2}}}{\left|
E(t)\right| ^{1/2}}\Bigl(\frac 52\frac 1{\mvir(t)}-
\frac{d\left| E(t)\right| }{2\left| E(t)\right|d\mvir(t)}\Bigr),
\end{equation}
where $E$ is the energy 
of the halo at time $t$, and $\lambda$ is the halo spin
parameter assumed here to be constant in time. The  contraction of 
the dark halo due to the gravitational drag of the infalling gas
is calculated within an adiabatic invariance scheme. A nearly exponential 
disc surface density arises naturally in this inside-out 
scheme of disc formation (e.g. FA00). The  
$\lambda$ parameter determines the scale length and surface density of the disc. 
According to analytical and numerical studies, $\lambda$ has a log-normal 
distribution with an average of $\sim0.05$ and a 
width in the logarithm of less than $1.0$ (e.g., Catelan \& Theuns 1996 
and references therein).

\begin{figure}
\vspace{12.6cm}
\includegraphics{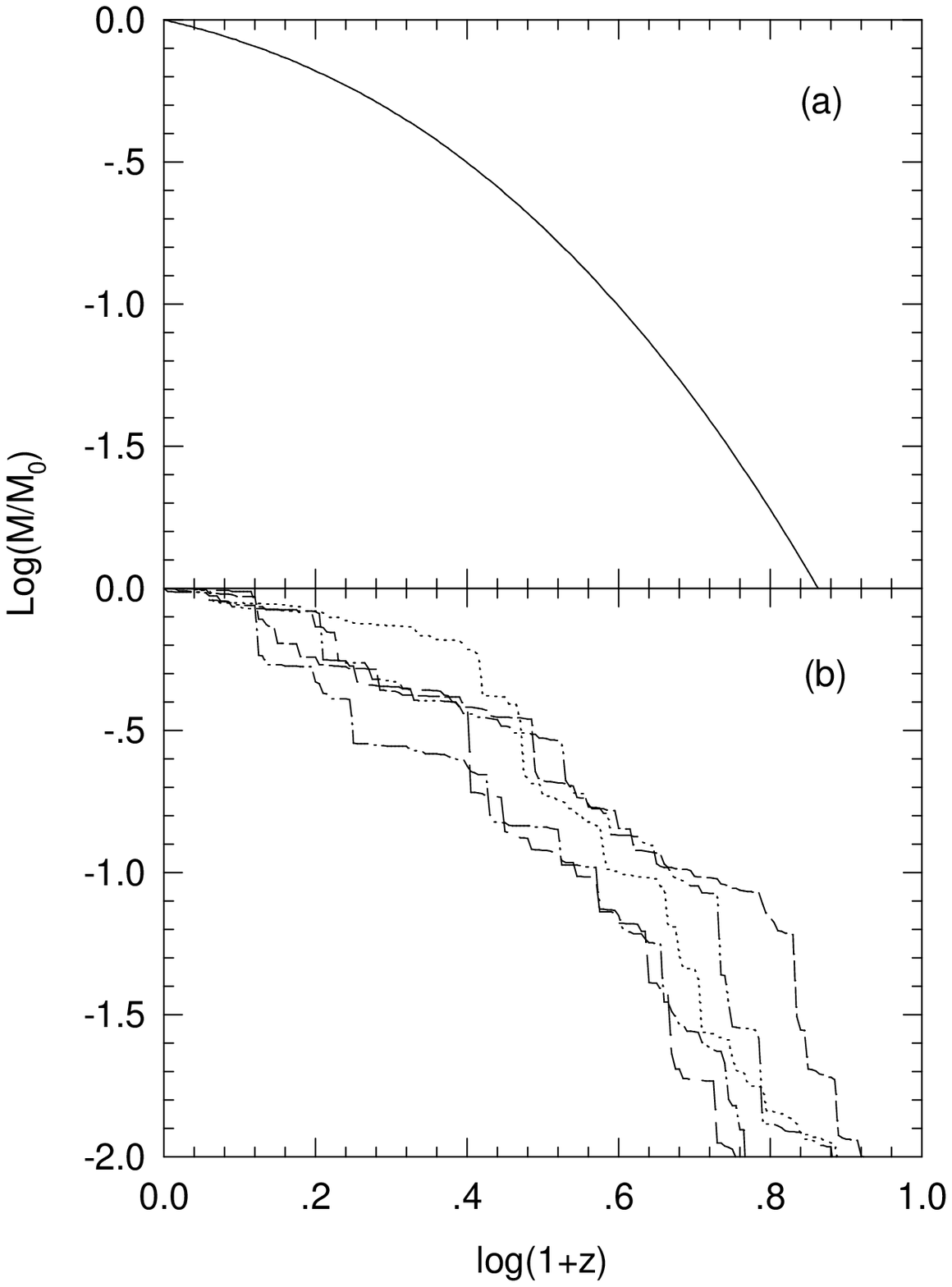}
    \caption{a) Average MAH for a galaxy having the present day asymptotic 
circular velocity, V$_{50}$, as the Milky Way. b) Five random realizations of 
cosmological MAHs yielding the same final asymptotic circular velocity as in a).}
\end{figure}

We note that if the dynamical consequences
of the MAHs we are using were taken into account, 
the repeated merging of substructures would destroy the forming discs
(Navarro \& Steinmitz 2000). This is a shortcoming general to all hierarchical
galactic formation models, related to the overabundance of substructure
seen at galactic scales (Klypin et al. 1999, Moore et al. 1999a), 
the excessive mass concentration in the central region of CDM haloes 
(e.g., Moore 1994, Burkert \& Silk 1997, Hernandez \& Gilmore 1998), and the 
loss of angular momentum of the gas during incorporation into the forming galaxy 
(Navarro \& Benz 1991). All these difficulties might indeed imply modifications 
to the overall standard scenario.
At this point we introduce as a working hypothesis the assumption that 
such modifications might result only in a more gentle and homogeneous mass accretion, 
without altering the overall structure formation and growth model, which
appears at the moment a very robust prediction of the Gaussian initial fluctuation
hypothesis within the inflationary CDM cosmology.

\subsection{Disc star formation}

As the disc begins to form in the centre of the growing 
CDM halo, transformation of cold gas into stars takes place
whenever Toomre's instability parameter for the gaseous disc, 
$Q_{g} (r)=v_g(r)\kappa(r)/\Sg(r)$, is less than a given threshold, $q$ 
(where $v_g$ and $\Sg$ are the gas turbulent rms velocity 
and surface density, and $\kappa$ is the epicyclic frequency, Toomre 1964). 
A fraction of the formed stars explodes 
as SNae, reheating the gas disc column and increasing $Q_g(r)$ above the
instability threshold. This inhibits SF temporarily, 
whilst the gas disc column dissipates its turbulent energy, 
decreasing $v_g$ back to the value when $Q_g(r)$ becomes smaller
than $q$. Actually we calculate the SF rate (SFR) directly from the 
equation that relates the energy 
input with the energy dissipation assuming that $Q_g (r)$ is always 
equal to $q$ at all radii, i.e. we allow only for the stationary solution (a more 
detailed description can be found in Firmani \& Tutukov 1994; Firmani et al. 1996).
We also consider heating due to the infalling gas, 
although its contribution is much smaller than the SN input.    

While the SFR is rather insensitive to $q$, the gas disc height
strongly depends upon it. Numerical and observational studies suggest thresholds of the
order of $q=2$ (see references in FA00). Further, for $q=2$,
we obtain realistic values for the gas disc height of a Milky Way (MW) model 
at the solar neighbourhood. The turbulent dissipation time $t_d(r)$ 
is calculated following a simple approach given in Firmani et al. (1996):
\begin{equation}
t_d=\alpha \  \frac{\delta r}{v_g}= \alpha \ \frac{1}{V/r+dV/dr}= \alpha
\ \frac{2\Omega}{\kappa^2}
\end{equation}
where $\delta r$ is the mean free radial distance traveled by a turbulent cell
at Galactocentric radius $r$ with a speed $v_g$ ($\delta r$ is assumed 
to be much smaller than the orbital length), $V$ and $\Omega$ are 
the circular and angular 
velocities, and $\alpha$ is a parameter close to unity. Estimates of 
the turbulent dissipation time in the solar 
neighbourhood obtained in compressible magneto-hydrodynamic 
simulations of the ISM (Avila-Reese \& V\'azquez-Semadeni 2001) give a 
lower limit that is roughly half the time given by Firmani et al. (i.e. 
$\alpha\grtsim 0.5$ in eq. (2)). It should also be taken into account 
that in the 
real ISM, besides turbulence, magnetic fields and cosmic rays contribute 
to the pressure, in such a way that the $t_d$ found in the turbulence 
simulations could be only a lower limit to the global energetic dissipation 
time in the ISM. We shall use here $\alpha =0.7$   

The evolution of the stellar populations is followed with
a parameterization of simple population synthesis models (see Firmani
\& Tutukov 1992, 1994). A Salpeter IMF with minimal and maximal masses
of 0.1\msun\  and 100\msun, and solar metallicities were used. 

The azimuthally averaged dynamics of the 
evolving gas and stellar discs coupled with the dark halo are 
treated by solving the corresponding hydrodynamical equations. 
With our method we obtain the temporal evolution 
of the gas infall rate on the disc $\infrate (r,t)$,  
the SFR $\sfr(r,t)$,
the disc gas and star surface densities, $\Sg(r,t)$ and $\ssd(r,t)$, and the circular 
velocities due to the disc and halo, V$_{\rm d}(r,t)$ and 
V$_{\rm h}(r,t)$, respectively at all radii. In our model the halo structure and 
the disc gas infall rate are directly related to the 
cosmology.

We find that the local SFR typically follows a Schmidt law, 
$\sfr (r)\propto \Sg^n$, with $n\approx2$ along a major portion of the 
disc (Firmani et al. 1996; AF00). This SF law is basically 
a consequence of the self-regulation mechanism we use. 
It is important to mention that in 
our models, self-regulation takes place only {\it within the disc}
and not at the level of a hypothetical intra-halo medium
in virial equilibrium with the CDM halo (e.g., White \& Frenk
1991; Cole et al. 1994; Baugh et al. 1997; van den Bosch 1998;
Somerville \& Primack 1999). The ISM of normal
disc galaxies is dense and very dissipative. Hence, one expects 
gas and kinetic energy outflows to be confined mostly
near the disc (see numerical results and more references in 
Avila-Reese \& V\'azquez-Semadeni 2001). Nonetheless, in some
cases and at early epochs, the disc-halo feedback may be important;
in our model this ingredient may be taken into account by the free parameter 
\fd, at least in a first approximation.

\begin{table}

  \caption{Properties of the MW: observations and model results}
\label{tbl-1}
 \begin{tabular}{|l|r|r|r|} \\ \hline \\ \hline
Observable & Predicted value $^{a}$ & Observed value & Reference \\ 
\hline \\ \hline
$V_c(50)$ $^{b}$ [\kms] & 208 & $ 206 \pm 10$ & 1,2 \\
\rs [kpc] & 3.0  & 3.0 $\pm$ 0.5 & 3 \\
M$_{\rm s}$ $^{c}$ [$10^{10}$ \msun] & 4.4 & 4-5 & 4,5\\ \hline
$\mvir$ [$10^{12}$ \msun] &  2.8  & 1-4 & 2,6 \\
$\vm^{d}$ [\kms] & 235  & 220 $\pm$ 10 & 7,8,9,5 \\
$L_B$ [$10^{10} L_{B_{\odot}}$] & 1.7 & 1.8 $\pm$ 0.3 & 10,9\\ 
r$_{B}$ $^{e}$ [kpc] & 4.3 & 4-5 & 11\\
$\Sigma_{\rm 0,K}$ $^{f}$ [$L_{K\odot} pc^{-2}$] & 810 & $1000 \pm 200$ & 11\\   
\fg         & 0.23 & $0.15-0.20$ & 12 \\
SFR [\msun yr$^{-1}$] & 2.9 & 2-6 & 13,14 \\ \hline
{\it Solar neighborhood} \\ \hline
\ssd [\msun pc$^{-2}$] & 41.3 & 43 $\pm$ 5 $^{g}$ & 4  \\
\ssg  [\msun pc$^{-2}$] & 13.6  & 13 $\pm$ 3 & 15 \\
\sfr [\msun Gyr$^{-1}$ pc$^{-2}$] & 3.1 & 3-5 & 16,17 \\ 
$B-K$ [mag] & 3.15 & 3.13 & 8 \\ \hline
\end{tabular}

\begin{flushleft}
$^{a}$ {The MW model was obtained tuning three input parameters in order
to reproduce the first three quantities, which are constraints and not
predictions}\\
$^{b}$ {Circular (asymptotic) velocity at 50 kpc.}\\
$^{c}$ {Stellar (disc+bulge) mass.}\\
$^{d}$ {This maximum rotation velocity \vm\ does not take into account
the nuclear and bulge region and it is for Galactocentric radii smaller than 15 kpc;
the observational determinations at larger radii are uncertain.}\\
$^{e}$ {$B-$band disc scale length.}\\
$^{f}$ {Disc central surface brightness in the $K-$band, taking a mass to light 
ratio of 1.0 for this band.}\\
$^{g}$ {This estimate includes the contribution of stellar remnants.}\\
\vskip 5pt
\item References: 1. Kochanek 1996; 2. Wilkinson \& Evans 1999; 
3. Sackett 1997; 4. Mera et al. 1998; 5. Dehnen \& Binney 1998a;
6. M\'endez et al. 1999; 7. Fich \& Tremaine 1991; 8. Binney \&
Merrifield 1998;
9.  Binney \& Tremaine 1987; 10. van der Kruit 1986; 11. Kent et al. 1991; 
12. see BP99;
13. Prantzos \& Aubert 1995; 14. Pagel 1997; 15. Kulkarni \& Heiles 1987; 
16. Talbot 1980
17. Rana 1987

\end{flushleft}

\end{table}

\section{Modeling the Milky Way}

For a given cosmology a model is completely determined by the virial 
mass \mvir, the MAH, the spin parameter $\lambda$, and the disc mass fraction \fd, 
as all other parameters are fixed by physical considerations.

In order to calculate a model representative of the MW, we have to 
correctly chose these input parameters so as to fit the observational constraints.
Fortunately, each one of these parameters is tightly related to a given 
present-day MW feature, although with some 
(small) degeneracy. Due to this degeneracy and to the 
observational uncertainties in the MW observational constraints, 
it is necessary to carry out some ``parameter-space'' exploration.

The total mass of the MW is amongst the 
most poorly known of all Galactic parameters. Several estimates give values 
around $1.0-4.0\times 10^{12} \msun$ (see e.g., Wilkinson \& Evans 
1999 and more references therein). A more robust quantity is the 
mass within a certain large radius (50 kpc for example) constrained by the motions
of satellite galaxies, globular clusters, the local escape velocity
of stars, etc. Using these constraints together with the observed
rotation curve of the disc, Kochanek (1996) determined the MW mass
within 50 kpc. This mass implies a circular velocity V$_{50}$ of 
$206^{+10}_{-11}$ \kms ($206^{+22}_{-25}$ \kms, at $90\%$ C.L.).
A more recent determination of $V_{50}$ by Wilkinson \& Evans (1999),
constructed from the current data set of objects (27) 
with known distances and radial velocities at Galactocentric radii greater 
than 20 kpc, is consistent with the values found by Kochanek (1996).
A V$_{50}$ of $206^{+10}_{-11}$ \kms is the observational constrain 
we use to fix  the virial mass of the MW model for a given MAH.
Since the MAH determines the halo density
profile, \mvir\ will be different for different MAHs; for the average MAH, this
mass is $\mvir =2.8\times 10^{12}$ \msun.

The spin parameter $\lambda$ mainly influences the stellar disc 
scale length and the surface density. The stellar scale length \rs\ of the MW 
inferred from observations in infrared bands or estimated
from dynamical constraints is typically $3.0\pm 0.5$ kpc (Sackett 1997),
this becomes our second constraint. 
However, see for example Drimmel \& Spergel 2001 where values even smaller than these were found. 
For the average MAH, the observed central value
of \rs\ yields $\lambda=0.02$.

Once \mvir\ is given, \fd $\equiv \md/\mvir$ is fixed by the MW 
disc+bulge mass \md. Most baryons are
in the disc+bulge system: $4-5\times 10^{10}$ \msun\ in form
of stars (\ms, our third constraint) and at least $\sim 0.6-1.0 
\times 10^{10}$ \msun\ in form of gas
(M$_{\rm g}$) (see Table 1). The estimated masses of the stellar halo ($\approx
4.6 \times 10^{9}$ \msun, Haud \& Einasto 1989) and of all MW satellite
galaxies ($\sim 6-9 \times 10^{9}$ \msun) represent only a minor contribution 
with respect to the disc+bulge system. Taking the average MAH we obtain \fd =0.021.

The influence of different MAHs is felt upon the integral galaxy color 
index, gas fraction \fg, and scatter in the Tully-Fisher relation. 
Unfortunately the MAH cannot be described by one-parameter and the color
index and gas fraction are not accurately determined for the MW. 
Thus, we are left with a large range of possible MAHs, all of which lead to 
models with the same $V_{50}$, \rs\ and \ms\ for the MW.
However, the SFH in the disc of the evolving galaxy, and particularly at the solar
radius, is always highly sensitive to the input MAH. The range of MAHs allowed by
the structural constraints taken hence defines a range of possible SFHs for the
solar neighbourhood, with the most likely being that associated to the average MAH.
A comparison with the inferred SFH of the solar neighbourhood, explored in \S 4,
in fact shows the MAH of our Galaxy to have been close to the average one for 
systems of the MW mass.

\begin{figure}
\vspace{11.4cm}
\includegraphics{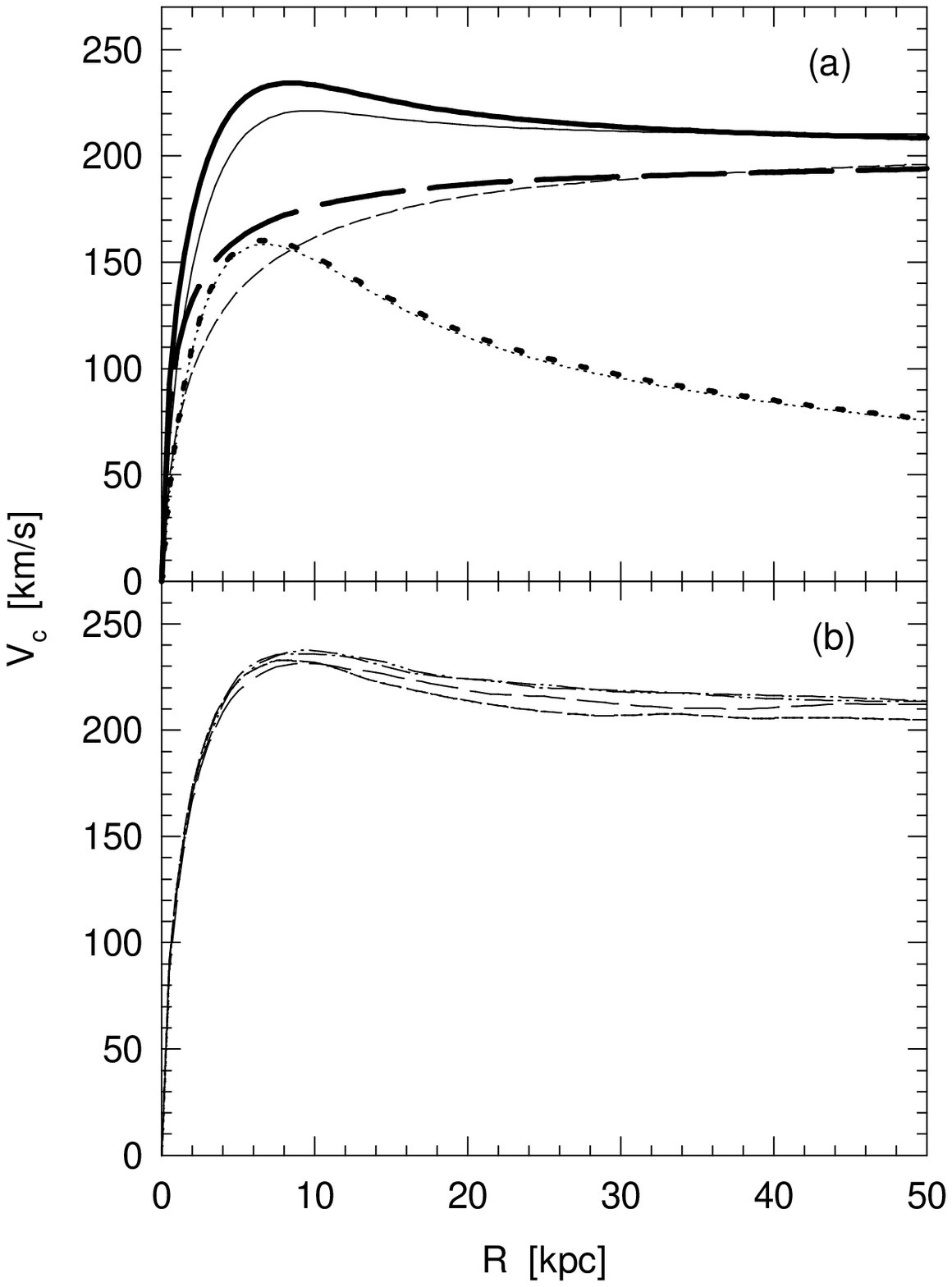}
\caption{a) Rotation curve decomposition for the fiducial Galaxy model giving the total,
halo and disc contributions, solid dashed and dotted thick lines, respectively. 
The thin lines are analogous to the thick ones, showing the case of artificially 
imposing a 10 kpc constant density core on the dark halo. 
b) Different final rotation curves, for MW models resulting from the various 
MAHs shown in Fig. 1b, with line style corresponding to the particular cases 
of Fig. 1b.}
\end{figure}

The observational estimates of V$_{50}$, \rs\ and \ms\ given in Table 1
are thus the constraints we require in order for a model to be 
considered representative of the MW.
For a given MAH the previous conditions allow to determine \mvir, $\lambda$ and \fd.
When the average MAH is assumed we obtain \mvir$=2.8 \times 10^{12}M_{\odot}$, 
$\lambda=0.02$ and \fd=0.021. These conditions determine our MW central
model, which represents a cosmological maximum likelihood solution, given the
structural parameters of our Galaxy (fiducial model). 

It should be noted that the values of \fd\ and $\lambda$ used in order
to reproduce the stellar mass and scale length of the MW, are smaller than
those used in other studies (Dalcanton, Spergel \& Summers 1997; Mo, 
Mao \& White 1998; FA00, AF00). The disc mass fraction in haloes, 
\fd, used in these works was $\approx 0.05$, only slightly smaller than 
the baryon to total mass fraction, $\Omega_{b} /\Omega_{m}$, 
for the conservative value of $\Omega_b h^2\approx 0.010$ (Hogan 1998)
and $\Omega_{m}=0.3$. This disc mass fraction increases to
$\fb\approx 0.15$ taking into account the higher
estimate of $\Omega_b h^2$ ($\approx $0.019) inferred from the 
deuterium absorption in quasars (Burles et al. 1998). Thus, for the
MW the baryon fraction ($\md+M_{\rm stellar \ halo}+M_{\rm satellites}+...)/\mvir$) 
seems to be smaller 
than the universal baryon fraction. Cosmological numerical simulations indeed
show that most of the baryons in the past and in the present are not 
within gravitationally bound systems, but in form of diffuse
and warm-hot intergalactic gas (Cen \& Ostriker 1999; Dav\'{e}t al. 2000).
Smaller values of \fd\ allow for smaller values of $\lambda$ 
(given a fixed \rs) that
otherwise could lead the disc to destructive gravitational instabilities
(see Mo et al. 1998; FA00). If the
distribution of $\lambda$ is mainly responsible for the galaxy distribution
in surface brightness, then, that high-surface brightness MW-type galaxies
should have $\lambda\approx 0.02$ agrees with a distribution where low
surface brightness (LSB) galaxies ($\lambda \grtsim 0.05$) are equally abundant (e.g.
McGaugh 1996). To conclude, the values for $\fd$ and $\lambda$
inferred for the average MAH seem to be consistent with several observational
and theoretical pieces of evidence.

\subsection{Rotation curve decomposition}

In Fig. 2a the rotation curve decomposition of the fiducial MW model is
shown by the thick lines, with the continuous curve giving the total
rotation curve, the dashed one the halo component, and the dotted one the disc 
contribution, at the present epoch.
The maximum of the final rotation curve occurs at at $\approx 2.2$ \rs\ with a value 
of 235 \kms.

The model rotation curve is somewhat more peaked than the observed one 
(Fich \& Tremaine 1991) and the 
halo component dominates over the disc component at almost all radii. 
This is a direct consequence of the high central mass concentration of 
the CDM halo. There is no way of obtaining a flat rotation curve for the
parameters of the MW when a CDM halo is used. The density profile
of our ``virgin'' CDM halo (without the gravitational drag due to disc 
formation) is well described by the NFW profile. 
Recent very high-resolution simulations show that the inner
density profile of CDM haloes is even steeper than the $r^{-1}$ slope
of NFW profiles (Moore et al. 1999b; see also Fukushige \& Makino 1997). 
The rotation curve of a MW model in this case would be even more 
peaked than that shown in Fig. 2, which should be taken as a
lower limit in terms of the central overshooting. The intrinsic
scatter in the density profiles of CDM haloes does not help to alleviate
this problem, as can be seen in Fig. 2b, where the total final
rotation curves corresponding to the five particular MAHs of Fig. 1b are shown.

\begin{figure}
\vspace{11.4cm}
\includegraphics{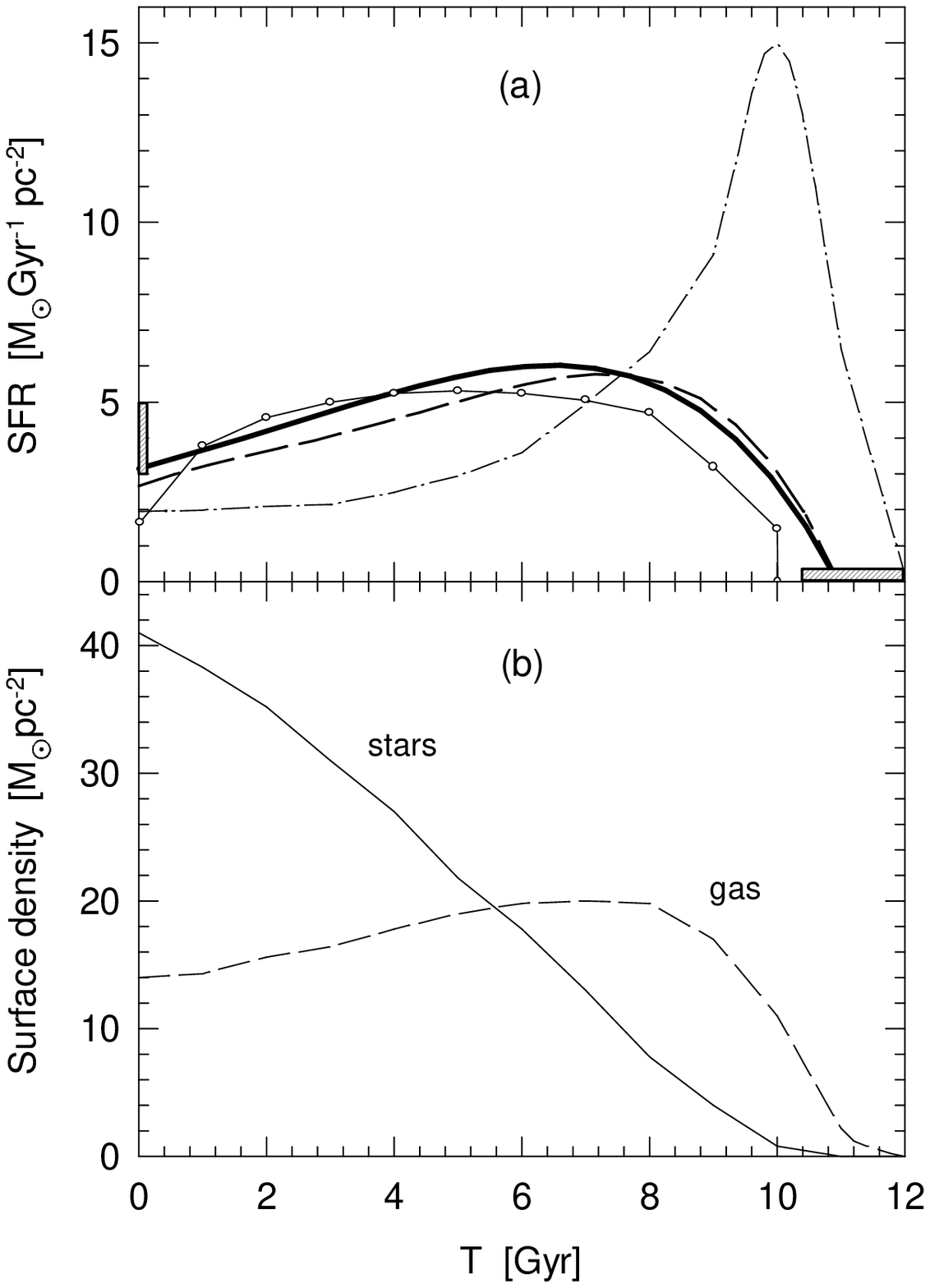}
\caption{a) The thick solid curve gives the predicted SFH for the solar 
neighbourhood corresponding to the most probable cosmological MAH (the average 
MAH), with the corresponding gas accretion history at the solar neighbourhood 
given by the dot-dashed curve. Also shown are the observational inferences of
BDB00 for the age of the solar neighbourhood, horizontal shaded box, and of 
Talbot 1980 and Rana 1987 for the present day value of the SFR, vertical 
shaded box. The jointed circles give the indicative inferences of Bertelli 
\& Nasi (2001). Our prediction for the average MAH clearly complies with the 
available observational restrictions for the local SFH.
b) This panel shows the evolution of the gas and stars surface densities at 
the solar neighbourhood, dashed and solid lines, respectively. The present day 
values being in good accordance with observations.}
\end{figure}

Several pieces of evidence point out that dark matter haloes at all scales indeed
have constant density cores (Moore 1994, Flores \& Primack 1994 and Burkert 1995
in dwarf galaxies, 
de Blok \& McGaugh 1997 in LSB galaxies, Hernandez \& Gilmore 1998 and Salucci 2001 
for normal spirals, Tyson, Kochanski \& Dell'Antonio 1998 for galaxy clusters). 
According to the analysis of rotation curves of dwarf and LSB 
galaxies carried out in Firmani et al. (2001), the core radius of the 
``virgin'' halo of a galaxy of the size of the MW should be 
$\approx 10$ kpc with a central density of $\approx 0.01-0.02$ 
\msun pc$^{-3}$. To illustrate the effect that such a constant density core
would have in terms of the rotation curve of the MW, we artificially smooth the 
inner density profile of 
the MW dark halo model along its evolution in such a way that at the 
present epoch it has such a core. The final rotation curve 
decomposition obtained after the formation of a MW disc in this halo is 
shown in Fig. 2a by the thin curves, which are analogous to the
thick curves of the NFW case. The rotation curve is now almost completely flat
and the disc component dominates over the halo component up to the
solar radius, i.e., the dashed and dotted curves now intersect at approximately
the solar radius, as direct studies of our galaxy show (e.g. Kuijken \& Gilmore 1989). 
We can hence conclude that a comparison of dynamical studies in the MW 
and galactic evolutionary models offers yet another evidence of the excessive 
central concentration of CDM haloes predicted by the standard cosmology. 
Fortunately, the inclusion of a shallow core in the halo has no
major consequences upon the SFH, as will be shown below.

\section{Star formation history in the solar neighbourhood}

As it is seen in Table 1, the properties of the MW fiducial model 
predict reasonably well the data inferred 
from observations, with the exception of the rotation curve.
In Fig. 3a we show the evolution of the SFR per 
unit area and of the gas infall rate, both 
at the radius $R_0=8.5$ kpc, $\sfr (R_0,t)$ (solid line) 
and $\infrate (R_0,t)$ (point-dashed line), respectively. The disc at 
the radius $R_0$ begins to form at a look-back time $\sim 11$ Gyr ($z\approx 2$).

An interesting test derives from taking into account an integrated colour index for the
solar neighbourhood. 
Using recent Padova stellar evolutionary models (e.g. Girardi et al. 1996) we 
computed synthetic colour magnitude diagrams for the average 
SFH, shown by the solid curve in Fig. 3a, taking a 
solar metallicity for the last 2 Gyr, and one third solar before this 
point, as a first approximation to the enrichment history. Calculating the $B-K$ 
colour index for realizations containing upwards of 200,000 stars we obtain a value 
of $B-K=3.15$, in excellent agreement with the solar neighbourhood observational estimate for this
quantity of $B-K=3.13$ (Binney \& Merrifield 1999). Repeating the experiment for 
SFHs resulting from MAHs deviating from the average one, we obtain practically 
identical values for the $K-$band luminosity, reflecting equal integrals under the 
SFH curves, but rather different values for the $B-$band luminosity, reflecting 
different recent SFHs. Taking the MAHs which differ most from the average (of the 
ones given in Fig. 1b), we obtain differences of 0.3 mag in the
predicted $B-K$ of the solar neighbourhood, rather larger than the observational 
uncertainties of the measured value. This important result leads us to expect the 
average MAH as the optimal choice for the Galaxy.

The recent observational 
inferences of Binney, Dehen \& Bertelli (2000 henceforth DBD00) use kinematical
data from the Hipparcos catalogue to break the age metallicity degeneracy at the
oldest turn off, identified also from the Hipparcos satellite by combining
photometric data with theoretical isochrones. Their results for the age of the
solar neighbourhood are shown by the shaded bar in Fig. 3a, and are clearly in good
agreement with our predictions for the average MAH.
The maximum rate of gas accretion at $R_0$ is attained at $\approx
10.5$ Gyrs, and strongly decreases after 
this epoch. The SFH attains a broad maximum at 8-6 Gyrs ($z=0.9 - 0.6$) 
and then smoothly declines by a factor of $\approx 2$ towards the
present day. The level of 3.1 \msun pc$^{-2}$ Gyr$^{-1}$ we obtain for this
quantity lies well within the range determined observationally, shown by the 
vertical shaded box at t=0 Gyr (Talbot 1980, Rana 1987).
A similar SFH is obtained for the MW model 
with a shallow core (thick dashed line), which illustrates the
robustness of our results regarding SFHs, 
to the existing discrepancies in dark halo density profiles
between theory and observations.  

\begin{figure}
\vspace{7.7cm} 
\includegraphics{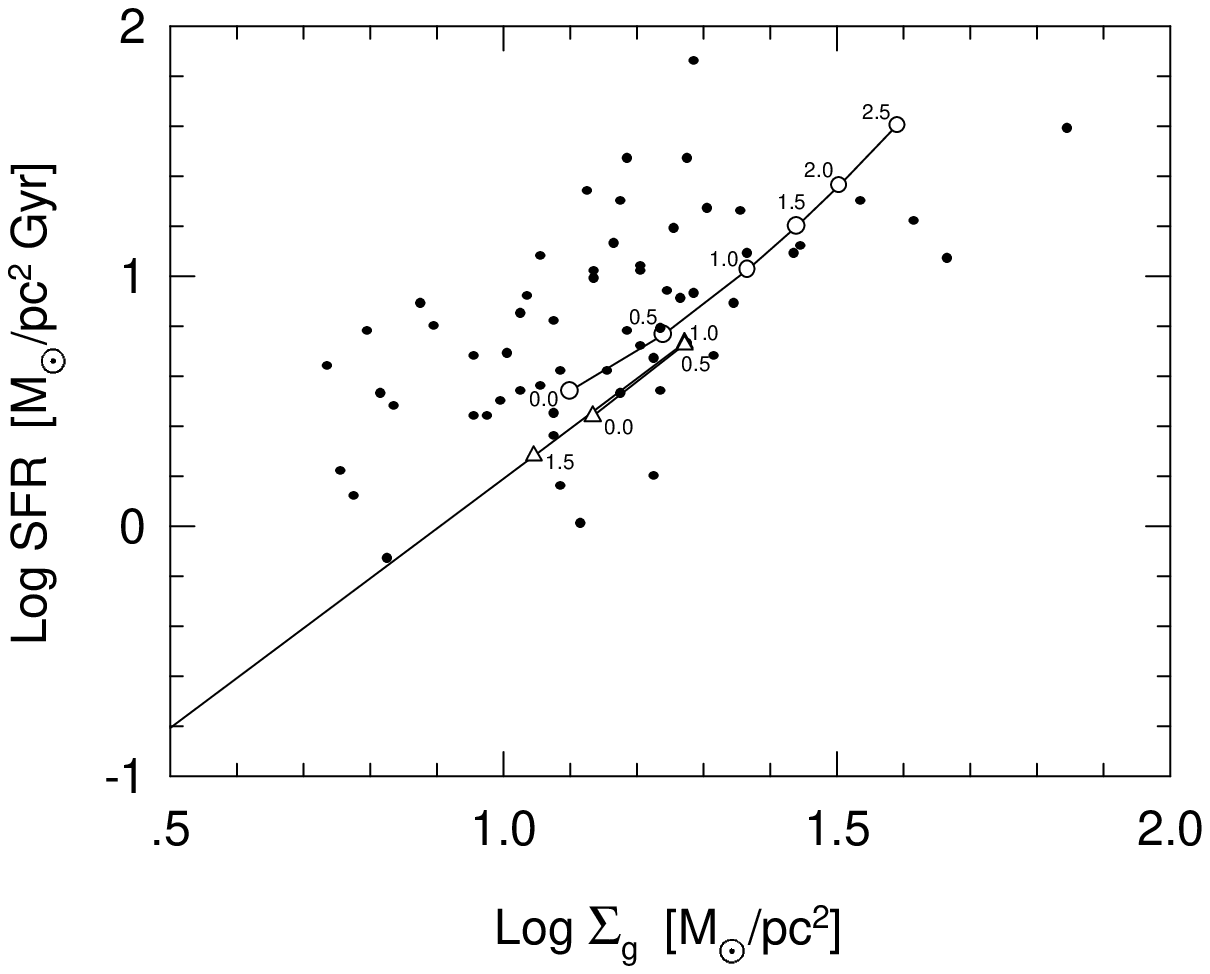}
\caption{Evolution of the relation between surface SFR and
surface gas densities, at the given redshifts, for the solar neighbourhood 
and the integrated Galaxy, jointed triangles and circles, respectively. A 
Schmidt law of power $\approx 2$ is clearly seen as the final result of the 
SF physics we used. The solid dots show the data of Kennicutt (1998), 
corrected to take helium into account, for 
which a Schmidt law of power $\sim 2$ is a good description.}
\end{figure}

Our SFR at the solar radius is almost indistinguishable from an average of an 
annulus with inner and outer radii of 7.0 and 10 kpc, respectively. The local 
SFH inferred from observations might refer to an annulus of this size 
because due to stellar diffusion, stars which are today in the solar 
neighbourhood could have been formed at Galactocentric radii differing by up
to $\sim 1.5$ kpc (e.g. Dehnen \& Binney 1998b). 

The difference in shape between the SFH, $\sfr (R_0,t)$, and the gas infall
history, $\infrate (R_0,t)$, depends on the turbulent dissipation time \td; the
shorter \td\ is, the closer $\sfr (R_0,t)$ is to $\infrate (R_0,t)$.
Nevertheless, the dissipation time we use is self-consistently calculated
according to Firmani et al. (1996),  and is in
rough agreement with results from MHD numerical simulations
of the ISM (Avila-Reese \& V\'azquez-Semadeni 2001).
We note that matching the constraints on the age of the solar neighbourhood
of BDB00 with a different value of $t_{d}$ would require the use of a MAH
differing largely from the average one, a rather improbable situation.

In Fig. 3a we also show \sfr\ for the solar neighbourhood, as inferred 
recently by Bertelli \& Nasi (2001), jointed circles. 
These authors used a volume-limited sample of 
field stars from the Hipparcos catalog and compared to 
synthetic color-magnitude diagrams generated for different 
parameter values of a SFH having a fixed parametric shape,  
 and different stellar initial mass functions. 
Applying a $\chi^2$ minimization, the best parameters of the {\it a priori}
given SFH are found. In order to pass from the SFR per unit
 volume to the SFR per unit area, the 
total number of stars of age $t$ found within the sampled sphere, as 
a fraction of the total over the disc height, a time dependent
correction factor, must be known. This 
can be obtained with a procedure which takes into account the 
details of the vertical disc force law and the variations of
velocity dispersion with age (e.g., Hern\'andez et al. 2000b). 

The SFH shown in Fig. 3a corresponds to the solution for 
the model with more degrees of freedom of the ones
used in Bertelli \& Nasi (2001) (their var-var model with $I_b=2$),
and including the volume to surface density correction mentioned above.
It must be noted that the detailed shape of the observed SFH is the result
of the age dependent correction factor, as well as of the volume SFH solved for.
As the authors emphasize, their method is useful to describe
only the {\it general trend of the SFH over the total life of the 
system}, and not its detailed shape.

As it can be concluded from the $\chi^2$ plots shown by these authors it is 
that the average SFR over the period 10-6 Gyr (in look-back time) could not 
have been larger than that over the 6-0 Gyr period. From Fig. 3a, this general 
trend is seen to agree well with the prediction of our fiducial model for 
the MW, suggesting that the main ingredients of the SF process in the MW were 
correctly taken into account, and that the MAH of the MW 
has been close to the average one. 

The integration of the SFR in time gives the stellar surface
density \ssd, shown by the solid line in Fig. 3b, 
as a function of time, at $R_0$. The dashed 
line is the gas surface density \ssg. At the present epoch, $\ssd(R_0)=
41.3 \ \msun$pc$^{-2}$ and $\ssg(R_0)= 13.6 \ \msun$pc$^{-2}$ in good 
agreement with observational measurements
(Table 1). The evolution of the stellar surface density at
the solar neighbourhood obtained with our model compares well to 
that derived by Fuchs et al. (2000) from a sample of G and K 
dwarfs with individually determined ages using their chromospheric 
HK emission fluxes, and assuming a given gas accretion history
(see their Fig. 7).

\begin{figure}
\vspace{11.4cm}
\includegraphics{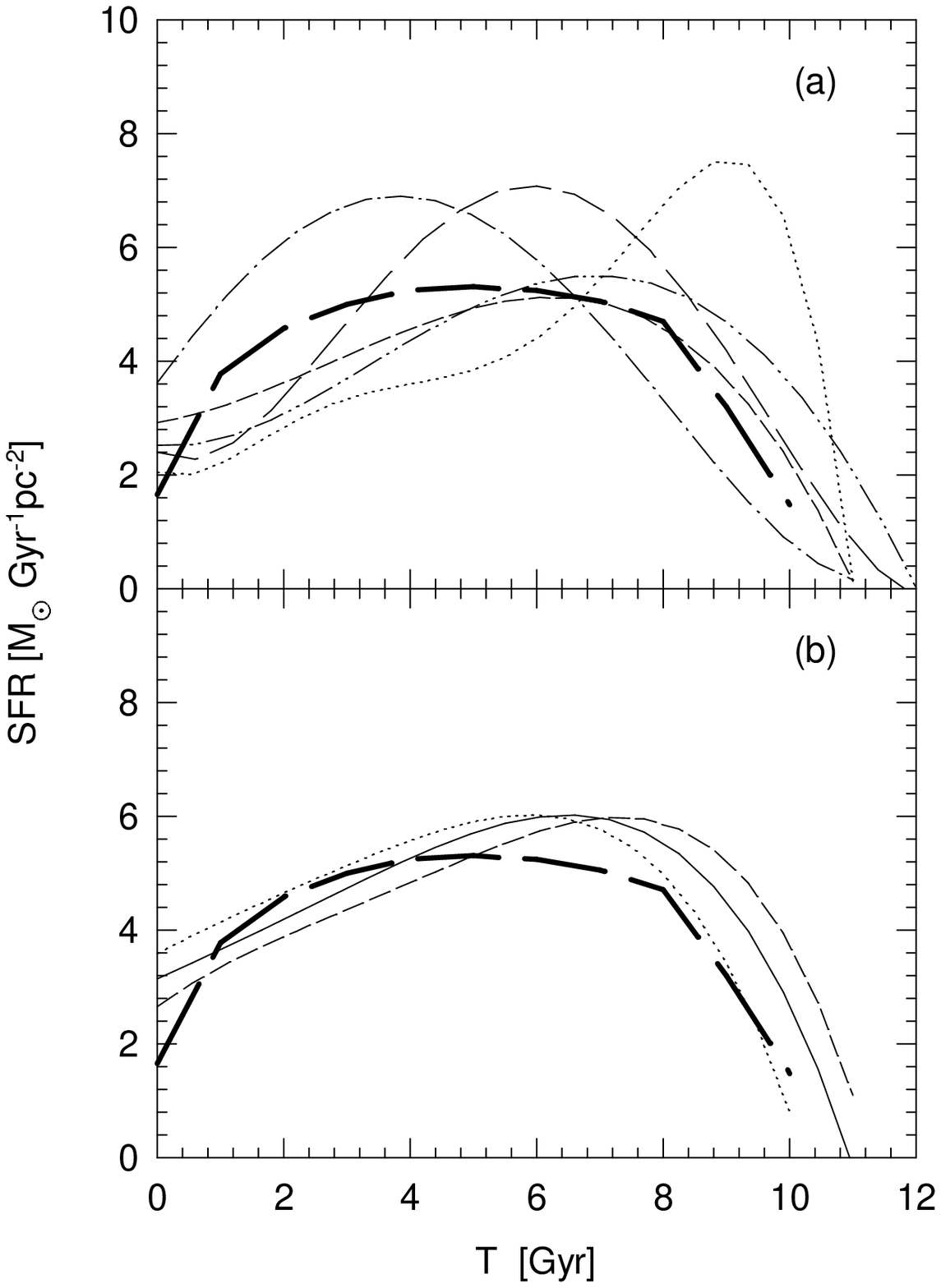}
\caption{a) SHFs resulting from the five random MAHs of Fig. 1b, in matching 
line styles. The thick dashed curve shows the indicative results of Bertelli 
\& Nasi (2001). b) Here we illustrate the variations in the predicted SFH of 
the solar neighbourhood as resulting from changing the input value of $\lambda$ 
in consistency with the uncertainties in the observational determinations of the 
disc scale length, no strong dependence is seen. The thick dashed curve is the 
same as in panel a).}
\end{figure}

Fig. 4 presents a global check of our SF mechanism. The open triangles 
show the time evolution of the solar neighbourhood in a log SFR vs. 
log \ssg\ plot, at z=0, 0.5, 1 and 1.5. The open circles show the temporal 
evolution for the whole modeled galaxy at z=0, 0.5, 1, 1.5, 2 and 2.5.
The solid dots show the observations of normal galaxies by Kennicutt (1998)
corrected to take helium into account. 
We have not included here the ULIRGs of that sample because 
they are not representative of the SF regime of normal galaxies but are related
to very strong SF bursts. In these cases, when the SFR exceeds the critical value
$\dot{\Sigma}_{\rm s,c}=300\msun$pc$^{-2}$Gyr$^{-1}$ ($\Sigma_{\rm
g,c}=100\msun$pc$^{-2}$) then SN energy injection no longer controls the SF,
the control is transferred to  the radiation pressure (Eddington limit). This
duality in the SF self-regulation mechanism has been studied by Firmani \&
Tutukov (1994). For normal galaxies where the SFR is typically less than
$\dot{\Sigma}_{\rm s,c}$ (the case of this work) self-regulation by SN energy
injection leads to a  Schmidt law of power $\sim 2$, while for the bursting SF
regime with SFR greater than $\dot{\Sigma}_{\rm s,c}$ the self-regulation by
radiation pressure leads to a  Schmidt law of power $\sim 1$. 
The comparison of the open and filled symbols shows a rather satisfactory 
agreement indicative of a Schmidt law of power $\sim 2$. 
The slight offset of the data
towards higher values of the SFR is not particularly meaningful because of
the scatter of the observational data and the uncertainty on the turbulent 
dissipation time.

The shape of our resulting SFHs for the whole MW disc differ from that at 
the solar radius. The inside-out formation of the disc and the mechanism of 
SF used here indeed lead to a differential evolution with radius of the galaxy 
(see also Boissier \& Prantzos 1999, hereafter BP99), which highlights
the importance of a dynamical galactic model in studies of 
galactic evolution where comparisons are made with spatially
localized data.

\subsection{Sensitivity of the local SFH to the model input parameters}

The input parameters of the MW evolutionary model 
presented above were chosen in such a way that the present-day
properties obtained by the model agree with the observed structural
parameters of the Galaxy, $V_{50}$, \rs\ and $M_{s}$
(see Table 1). Since the observational data are uncertain to 
some level and the model presents some degeneracy with
respect to the input parameters, we need to explore
the sensitivity of the predicted SFH in the solar neighbourhood
to this input.

Due to the difficulty in determining the MAH for the MW from any
observational constrain (see \S 3), we have taken the average MAH (fiducial model)
as a first choice, as it represents the most probable cosmological solution
for a MW mass galaxy. This obviously presents a smooth shape (Fig. 2a).
However, other MAHs are also able to 
produce dark haloes consistent with the MW. In Fig. 1b we have five MAHs
taken at random which, for a proper choice of the input parameters,
can also give models in agreement with the main conditions we fixed 
for the MW.
These MAHs are intrinsically discontinuous due to the stochastic
nature of the primordial density fluctuation field. The SFHs at 
R$_0=8.5$ kpc for the MW models produced with these MAHs are 
plotted in Fig. 5a, with the line style corresponding to the
MAHs of Fig. 2a. We have included a smoothing on the results
of these SFHs to facilitate the comparison with the results of
Bertelli \& Nasi (2001) (thick curve), as those results have a low time resolution.

As it can be seen, both the age of the solar neighbourhood and its
present-day SFR are sensitive to the assumed MAH
of the Galaxy. It is therefore these dependences that allows us to
constrain the MAH of the MW. The observed age of the
solar neighbourhood and its present SFR imply that the MAH of our
Galaxy has been very close to the average one for present day objects
of that mass.

The spin parameter $\lambda$ is mainly constrained by the disc scale 
length \rs, which has an observational uncertainty of 1 kpc. 
In Fig. 5b we present the SFHs corresponding to
MW models with the average MAH but with \rs=2.5 kpc (dotted line) and
\rs=3.5 kpc (dashed line), the two extremes of the observational 
range. The difference in the SFHs is small. The more 
concentrated disc (smaller $\lambda$) results in a later formation of the
solar neighbourhood than the less concentrated one (larger $\lambda$). The 
maximum in the SFH thus shifts slightly in time for these two alternative
models, although much less than the changes resulting from varying
the MAH. Therefore, uncertainties in the scale length do not
alter our conclusions regarding the connection between the MAH and the SFH of 
the Galaxy.

Changes in \fd\ and \mvir\ allowed by the uncertainties in $V_{50}$ and \ms\
basically produce upward/downward shifts of the local  
SFH directly proportional to \ms, the shape of the SFH 
remaining almost the same. No large variations of these parameters are possible, 
if we want the present day \ssd\ and \ssg\ at the solar neighbourhood to remain 
within the observed ranges.

Finally, we note that the Hubble parameter $H_{0}$ plays a
role in determining the cosmic time when the disc starts
to form at the solar neighbourhood. The predicted SFH in the solar 
neighbourhood starts at lower look-back time when $H_{0}$ is larger.
For example, if we set $H_{0}=80 \kms$Mpc$^{-1}$, then the disc
at the solar neighbourhood forms at $\approx 9.0$ Gyr, for the average MAH.

We find that
the disc age at R$_0$ for the fiducial model with $H_{0}=65 \kms$Mpc$^{-1}$ is 
consistent with observational estimates of the age of the solar neighbourhood
(BDB00). Thus, the data offer a joint consistency check on the cosmological
scenario, the galactic evolution model and the SF physics used.

\section{Discussion}

\subsection{Comparison with chemo-spectrophotometric models}

In our approach, the local and global SF and infall gas histories 
of galaxies are predicted in a deductive way. In the case of models of 
chemical and spectrophotometric evolution, 
these histories are given in a parametric form and then  
constrained by comparing the model results with observations. 

Much as the inversion of the high quality Hipparcos data allows for
a reconstruction of the SFH of the solar neighbourhood,
the large amount of observational constraints both for the solar neighborhood
and over the whole disc allow the chemo-spectrophotometric models to constrict 
these histories (e.g., Prantzos 
\& Aubert 1995; Chiappini et al. 1997; Prantzos \& Silk 1998; BP99). 
However, most of the restrictions
being time-integral constraints, the temporal resolution attained is low.   
 
The SFHs in the solar neighborhood reported in Prantzos \& Silk (1998) and 
in BP99 declines smoothly to the present epoch by a factor around two
since 
its maximum, which is attained $\sim 8-9$ Gyr and $\sim 6-7$ Gyr ago,
respectively. The main constraints for this smooth behaviour of the SFH,
which agrees qualitatively with our prediction, are the observed 
differential metallicity distribution of long-lived G-dwarfs and 
the age-metallicity relation. These constraints also impose the necessity
of a gradual gas infall, which again is in qualitative agreement with predictions of 
our model. 
 
However, due to the phenomenological nature of the
chemo-spectrophotometric 
models, a detailed reconstruction of the SFH is difficult to attain, in
particular 
at early epochs. For example, in both papers, Prantzos \& Silk (1998) and
BP99, the solar neighborhood is assumed to form 13.5 Gyr ago, which is
too early 
according to accurate kinematical and photometric studies 
(BDB00). Nevertheless, with an appropriate combination 
of the {\it a priori} gas infall history and SF law, 
the corresponding metallicity and surface density constraints have been 
roughly satisfied in spite of the early formation of the solar
neighborhood (degeneracy).

In as much as our results for global and local SFHs and gas infall rates
do not deviate from those inferred by requiring consistency with the observational
restrictions described above (BP99), we can expect our models to be also in 
agreement with the large variety of data regarding metallicities and abundance 
gradients in our Galaxy and for the solar neighbourhood.

\subsection{Stationary and bursting star formation}

Although we have assumed throughout that the main trends of the SF processes in 
the disc are well described by the stationary approach adopted, we are
aware of the existence of ample evidence in favour of a more bursting
character for this crucial ingredient, under some circumstances. However,
as will be detailed below, we have reasons to believe that this alternative
channel of SF will only introduce fluctuations of short temporal and spatial
character, such that our stationary solution still represents the long term/large
spatial trends accurately.

The method applied by Bertelli \& Nasi (2001) allows a 
reconstruction of the broad trends in the local SFH over 
a long time interval ($\sim 10$ Gyr), but it is not useful to 
trace fluctuations of short duration. The question of significant
temporal fluctuations has been raised by Hernandez et al. (2000b)
who in a direct and statistically grounded study of the colour magnitude diagram 
of Hipparcos find a modulation of a factor of $\sim1.5$, with a period of
$\sim 0.5$ Gyr in the SFH of the solar neighbourhood. 
These authors interpret this as evidence of an enhancement
of the SF activity caused by encounters with the 
spiral arm pattern density waves, a physical feature which is absent from our models. 
It is precisely the slowly 
oscillating character of the SFH found in that work, that justifies
the interpretation of our results as the average over the inherent
fluctuations of the Galactic SF process, i.e. this defines the temporal
resolution of our present study to be of the order of $0.5$ Gyr

Even stronger time fluctuations in the SFH of the solar neighbourhood
have been suggested by  Rocha-Pinto et al. (2000), who argue against
any steady (or self-regulated) component of the local SF. These authors use 
an empirical chromospheric activity indices vs. age relations to affix an age to
each star in a sample of 552 late type dwarfs. However, the empirical relation
used by these authors presents a strong scatter of around a factor 2 in age
(their Fig. 22), which should be included as a time smoothing kernel on their
final results. Interestingly, if such a smoothing is applied on their SFH, 
a trend similar to that reported in Bertelli \& Nasi (2001) appears 
(see Fig. 8 in Rocha-Pinto et al. 2000). We therefore find the above study 
in fact strengthens any conclusions drawn from the work of Bertelli \& Nasi 
(2001), from which no objections to the SF physics we are using are apparent.

A further mechanism responsible for strong time variations
in the SF processes of galaxies is that of merger and
tidally induced SF enhancements. However, this applies mostly
to systems such as the strongly interacting galaxies
imaged by the $IRAS$ satellite (Firmani \& Tutukov 1994), or galaxies in 
dense cluster environments, see for a review Dultzin-Hacyan (1997).
It is clear from the late type and low density environment of
the MW that any alterations of the global SF due to interactions have been minor, 
and in all probability do not affect the conclusions of our study.

\section{Summary and Conclusions}

We have presented predictions of the SFH in the solar neighbourhood
making use of a galactic evolutionary approach, tightly linked to a
cosmological scenario. In this way we establish a direct connection between
the evolution of the solar neighbourhood and the cosmological background.
Within our deductive approach, galactic discs form inside out within growing
CDM haloes. Star formation is triggered by disc gravitational instabilities and 
is self regulated by an energy balance in the turbulent ISM. In previous papers 
we have shown the ability of this approach to predict exponential discs over 
several disc scale lengths, as well as to reproduce the Tully-Fisher relation
and several of the correlations across the disc Hubble sequence. 
Our results are also in qualitative agreement with the gas and SFR histories 
found to be consistent by comparing observations with chemo-spectrophotometric
models.

The allowed MAHs for the MW result in SFHs for the
solar neighbourhood which bracket the range of observational inferences.
The average MAH for objects having the present day
mass of the MW represents a cosmological maximum likelihood solution for 
the Galaxy. The SFH in the solar neighbourhood predicted for the above
possibility, in the flat $\Lambda$ CDM model used here, closely matches the
available observational data.

For the cosmological galaxy evolution approach presented, the
age of the solar neighbourhood as inferred from observations 
implies a delay in the conversion of the
accreted gas into stars, consistent with our physical formulation of the
SF process in disc galaxies.

For the value of $H_{0}=65 \kms$ Mpc$^{-1}$ used, the most probable MAH for 
our Galaxy implies an age for the solar neighbourhood of $\approx 11$ Gyr, 
in agreement with direct observational inferences.

The rotation curve decomposition of the MW
model is unrealistic if the disc forms within a CDM halo. If the
halo has a shallow core of size and central density as inferred
from rotation curves of dwarf and LSB galaxies, this
problem is solved. Since the SFH is not 
affected by this, our  conclusions
regarding it and the cosmological build up processes
of the Galaxy will remain for models where dark haloes 
present shallow cores.  

\section*{\bf ACKNOWLEDGMENTS}

The authors thank Liliana Hern\'{a}ndez, Gilberto Zavala and Carmelo Guzm\'an for
computing assistance. The work of V.A. was supported by CONACyT
grants J33776-E and 27752-E.


\begin{thebibliography}{}

\bibitem{}Aparicio A., Bertelli G., Chiosi C., Garcia-Pelayo J.M., 1990,
A\&A 240, 262
\bibitem{}Aparicio A., Gallart C., 1995, AJ, 110, 2105
\bibitem{}Avila-Reese V., \& Firmani C., 2000, Rev. Mex. Astron.
Astrofis.,  36, 23 (AF00)
\bibitem{}Avila-Reese V., \& Firmani C., 2001, in ``The VII Texas-Mexican
Conference on Astrophysics: flows, blows, and glows'', eds. W.H. Lee \&
S. Torres-Peimbert, Rev. Mex. Astron. Astrofis. Serie de Conferencias,
vol. 10, 97
\bibitem{}Avila-Reese V., \& V\'azquez-Semadeni E., 2000, ApJ, 553, 645
\bibitem{}Avila-Reese V., Firmani C., \& Hern\'{a}ndez X., 1998, ApJ, 505,
37 
\bibitem{}Avila-Reese V., Firmani C., Klypin A., Kravtsov A., 1999, 
MNRAS, 310, 527 
\bibitem{}Baugh C.M., Cole S., Frenk C.S., 1996, MNRAS, 283, 1361
\bibitem{}Bertelli G., \& Nasi E., 2001, AJ, 121, 1013.
\bibitem{}Binney J., Merrifield M., 1998, in ``Galactic Astronomy''
(Princeton University Press: New Jersey), p. 556
\bibitem{}Binney J., Tremaine S., 1987, in ``Galactic Dynamics'' 
(Princeton University Press: New Jersey)
\bibitem{}Binney J., Dehnen W., \& Bertelli G. 2000, MNRAS, 318, 658
\bibitem{} Buchalter A., Jim\'enez R., Kamionkowski M., 2001, MNRAS, 322, 43
\bibitem{}Boissier S., \& Prantzos N., 1999, MNRAS, 307, 857 (BP99) 
\bibitem{}Burles S., Nollett K.M., Truran J.N., Turner M.S., 1999, Phys.
Rev. Lett., 82, 4176
\bibitem{} Burkert A., 1995, ApJ, 477,  L25
\bibitem{}Carigi L. 1996, Rev. Mex. Astron. Astrofis., 32, 179
\bibitem{}Catelan P., \& Theuns, T. 1996, MNRAS, 282, 436
\bibitem{} Cen R., Ostriker J.P., 1999, ApJ, 519, L109
\bibitem{}Chiapini C., Matteucci F., \& Gratton R. 1997, ApJ, 477, 765
\bibitem{}Chiosi C., Bertelli G., Meylan G., Ortolani S., 1989, A\&A, 219,
167
\bibitem{} Cole S., Arag\'on-Salamanca A., Frenk C.S., Navarro J., \&
Zepf S. 1994, MNRAS, 271, 781
\bibitem{}Dalcanton J.J., Spergel D.N.,  Summers F.J., 1997, ApJ, 482, 659
\bibitem{} Dav\'e R., Cen R., Ostriker J.P., Bryan G.L., Hernquist L., 
 Katz N.,  Weinberg D.H., Norman M.L.,  O'Shea B., 2000, ApJ, submitted 
 (astro-ph/0007217)
\bibitem{} de Blok W.J.G., \& McGaugh S.S., 1997, MNRAS, 290, 533
\bibitem{} Dehnen W., Binney J., 1998a, MNRAS, 294, 429
\bibitem{} Dehnen W., Binney J., 1998b, MNRAS, 298, 387
\bibitem{} Drimmel R., Spergel D.N., 2001, ApJ, in press (astro-ph/0101259)
\bibitem{} Dultzin-Hacyan D., 1997, Rev. Mex .Astron. Astrofis. Serie de
Conferencias, vol. 6, 132
\bibitem{}Firmani C., Avila-Reese V. 2000, MNRAS, 315, 457 (FA00)
\bibitem{}Firmani C., Tutukov A.V., 1992, A\&A, 264, 37
\bibitem{}Firmani C.,  Tutukov A.V., 1994, A\&A, 288, 713
\bibitem{}Firmani C., Avila-Reese V., Hern\'{a}ndez, X., 1997, in 
``Dark and Visible Matter in Galaxies and Cosmological Implications'',  
ed. M.Persic \& P. Salucci,  ASP Conference Series  117, 424
\bibitem{}Firmani C., Hern\'{a}ndez X., Gallagher J., 1996, A\&A, 308,
403
\bibitem{}Firmani C., D'Onghia E., Chincarini G., Hern\'andez X., 
        Avila-Reese V., 2000, MNRAS, 321, 713 
\bibitem{} Fich M., Tremaine S., 1991, ARA\&A, 29, 409
\bibitem{}Flores R.,  Primack J.R., 1994, ApJ, 427, L1
\bibitem{}Fuchs B., Dettbarn C., Jahreis H., Wielen R. 2000, 
in Dynamics of Star Clusters and the Milky Way, ed. S. Deiters,
B. Fuchs, A. Just, R. Spurzem, \& R. Wielen, ASP Conference Series, in
press
\bibitem{}Fukushige T., Makino J., 1997, ApJ, 477, L9
\bibitem{}Girardi L., Bressan A., Chios C., Bertelli G., Nasi E., 1996, 
A\&AS, 104, 365
\bibitem{}Haud U., Einasto J., 1989, A\&A, 223, 95 
\bibitem{}Hern\'andez X., Gilmore G., 1998, MNRAS, 294, 595
\bibitem{}Hern\'andez X., Valls-Gabaud D., \& Gilmore G. 1999, MNRAS, 304,
705
\bibitem{}Hern\'andez X., Valls-Gabaud D., \& Gilmore G. 2000a, MNRAS,
317, 831
\bibitem{}Hern\'andez X., Valls-Gabaud D., \& Gilmore G. 2000b, MNRAS,
316, 605
\bibitem{}Hern\'andez X., Ferrara A., 2001, MNRAS, in press
\bibitem{}Hogan C.J., 1998, Space Science Reviews, v. 84, 127 
\bibitem{}Kauffmann G., Charlot S., Balogh M.L. 2001, MNRAS, submitted 
(astro-ph/0103130)
\bibitem{}Kauffmann G., White, S.D.M., \& Guiderdoni, B. 1993, MNRAS, 264,
201
\bibitem{}Kennicutt R.C. 1992, in Star Formation in Stellar Systems, 
ed. G. Tenorio-Tagle M. Prieto, \& F. S\'anchez (Cambridge
Univ. Press), 191
\bibitem{}Kennicutt R.C. 1998, ApJ, 498, 541
\bibitem{} Kent S., Dame T., Fazio G., 1991, ApJ, 378, 131
\bibitem{} Klypin A.A., Kravtsov A.V., Valenzuela O., Prada F., 1999, ApJ, 
        522, 82
\bibitem{}Kochanek C.S., 1996, ApJ, 457, 228
\bibitem{}Kuijken K., Gilmore G. 1989, MNRAS, 239, 571 
\bibitem{} Kulkarni S., Heiles C., 1987, in ``Interstellar Processes'', 
eds. D. Hollenbach \& H. Thronson (Kluwer), 87
\bibitem{}Lacey C.G., Cole S., 1993, MNRAS, 262, 627
\bibitem{}Lacey C.G., Silk J. 1991, ApJ, 381, 14 
\bibitem{} Lilly S. et al., 1998, ApJ, 500, 75
\bibitem{}Chiappini C., Matteucci F., Gratton R., 1997, ApJ, 477, 765
\bibitem{}Mao S., Mo H.J., White S.D.M., 1998, MNRAS 296, 847
\bibitem{}Madau P., Pozzetti L., Dickinson M., 1998, ApJ, 498, 106
\bibitem{}McGaugh S. S., 1996, MNRAS, 280, 337

\bibitem{} M\'endez R.A., Platais I., Girard T.M., Kozhurina-Platais V.,
van Altena W.F., 1999, ApJ, L39 
\bibitem{} Mera D., Chabrier G., Schaeffer R., 1998, A\&A, 330, 953 
\bibitem{}Mighell K.J., Butcher H.R., 1992, A\&A, 255, 26
\bibitem{}Mighell K.J., 1997, AJ, 114, 1458
\bibitem{} Mo H.J., Mao S.,  White S.D.M., 1998, MNRAS, 295, 319
\bibitem{}Mould J.R., Han M., Stetson P.B., 1997, ApJ, 483, L41
\bibitem{}Moore, B. 1994, Nature, 370, 629
\bibitem{}Moore, B., Ghigna, S., Governato, F., Lake, G., Quinn, T.,
Stadel, 
        J., \& Tozzi, P. 1999a, ApJ, 524, L19
\bibitem{}Moore B., Quinn T., Governato F., Stadel J., Lake G., 1999b, 
        MNRAS, 310, 1147
\bibitem{}Navarro J., Benz W., 1991, ApJ, 380, 320
\bibitem{}Navarro J., Steinmetz M., 2000, ApJ, 538, 477
\bibitem{}Navarro J., Frenk C.S., White S.D.M., 1997, ApJ, 490, 493
\bibitem{} Pagel B., 1997, in ``Nucleosynthesis and Galactic Chemical
Evolution `` (Cambridge: Cambridge University Press, 1997)
\bibitem{}Prantzos N., \& Aubert, O. 1995, A\&A, 302, 69
\bibitem{}Prantzos N., \& Silk, J. 1998, ApJ, 507, 229
\bibitem{} Rana N., 1987, A\&A, 184, 104
\bibitem{}Rocha-Pinto H., \& Maciel W. 1997, MNRAS, 325, 523
\bibitem{}Rocha-Pinto H., Maciel W.J., Scalo J., \& Flynn, C. 2000,
A\&A, 358, 869. 
\bibitem{}Sackett P.D., 1997, ApJ, 483, 103
\bibitem{}Salucci P., 2001, MNRAS, 320, L1
\bibitem{}Smecker-Hane T.A., Stetson P.B., Hesser J.E., Lehnert M.D., 1994
AJ, 108, 507
\bibitem{}Somerville R.S., \& Primack J.R. 1999, MNRAS, 310, 1087
\bibitem{}Struck C., \& Smith, D.C. 1999, ApJ, 527, 673
\bibitem{}Talbot R.J., 1980, ApJ, 235, 821
\bibitem{}Tolstoy E., Saha A., 1996, ApJ, 462, 672.
\bibitem{}Tyson J.,  Kochanski G.P., \& Dell'Antonio  I. P., 1998 ApJ,
498, L107
\bibitem{}van den Bosch F.C. 1998, ApJ, 507, 601
\bibitem{}van der Kruit P., 1986, A\&A, 157, 230
\bibitem{}White S.D.M, \& Frenk C.S. 1991, ApJ, 379, 52
\bibitem{}Wilkinson M.I., Evans W., 1999, MNRAS, 310, 645





\end{thebibliography}
\end{document}